\documentstyle[12pt,amssymb,epsfig]{article}
\author {Ernesto S. Loscar, Rodolfo A. Borzi and Ezequiel V. Albano$^{a}$\\
$^a${\it Instituto de Investigaciones Fisicoqu\'{\i}micas
Te\'{o}ricas y Aplicadas}\\{\it (INIFTA), UNLP, CONICET, 
Suc.4, CC16,}\\{\it
1900 La Plata, Argentina}}

\title{On the fluctuations of jamming coverage upon
random sequential adsorption on homogeneous and heterogeneous media.}
\begin{document}
\maketitle

\begin{abstract}

The fluctuations of the jamming coverage upon Random Sequential Adsorption (RSA) 
are studied using both analytical and numerical techniques.
Our main result shows that these fluctuations (characterized by $\sigma_{\theta_J}$)
decay with the lattice size according to the power-law
$\sigma_{\theta_J} \propto L^{-1/\nu}$. The exponent $\nu$ depends on the
dimensionality $D$ of the substrate
and the fractal dimension of the set where the RSA process
actually takes place ($d_{f}$) according
to $\nu = \frac{2}{2D - d_{f}}$.
This theoretical result is confirmed by means of extensive numerical simulations
applied to the RSA of dimers on homogeneous and stochastic fractal substrates.
Furthermore, our predictions are in excellent agreement with
different previous numerical results.

It is also shown that, studying correlated stochastic processes, one can
define various fluctuating quantities designed to capture either the
underlying physics of individual processes or that of the whole system.
So, subtle differences in the  definitions may lead to
dramatically different physical interpretations of the results.
Here, this statement is demonstrated for the case
of RSA of dimers on binary alloys.
\end{abstract}

\newpage
\section{Introduction}

Physical and chemical properties of adsorbed
monolayers are being studied with increasing interest
because their understanding is essential for the rationalization
of many phenomena and processes occurring on surfaces
and interfaces,
such as adsorption, desorption, catalysis, corrosion, wetting,
adhesion, diffusion, etc. The equilibrium behavior
of such overlayers can be described by
a Gibbs measure parametrized by the coverage $\theta$ and
the temperature $T$. Within this context, the critical
behavior of adsorbed films has extensively been studied \cite{6,7,8}.
On the other hand, numerous physical processes can be modeled
as the sequential, irreversible filling of a surface by atoms
or molecules. Some
examples are the reaction at specific sites on a polymer chain,
adsorption onto surfaces, reaction between groups on adjacent
surface sites, etc. \cite{noev,teno,Evans}.

Considering the irreversible deposition of particles on a surface,
one has two characteristic time scales: the time between depositions,
and the diffusion time of the particles on the surface. For very strong
interaction between particles and the substrate (chemical adsorption),
diffusion becomes irrelevant and the venerated
Random Sequential Adsorption (RSA) model provides an excellent
description of the underlying processes (for a review on
RSA models see e.g. \cite{Evans}).
Under these conditions the system evolves rapidly toward far-from
equilibrium conditions and the dynamics becomes essentially
dominated by geometrical exclusion effects between particles.
This kind of effects has been observed in numerous experiments \cite{9}.

When RSA involves adsorption on single sites, the case
is termed `monomer filling'. Also, processes involving adjacent pairs
of sites are referred as `dimer filling', while adsorption on larger
ensembles of sites corresponds to `animal filling' \cite{Evans}.
A quantity of central interest for the understanding of
RSA processes is the asymptotic value of the fraction $\theta$
of the total surface occupied by adsorbed objects, which is called the
jamming coverage $\theta_{J} = \theta(t \rightarrow \infty)$.
Within this context, the RSA of needles (or linear segments)
on homogeneous, two-dimensional samples, has very recently attracted
considerable interest \cite{Galam,Pekalski}. Particular
attention has been drawn to the interplay between jamming
and percolation \cite{Galam,Pekalski,frede}.
Of course, the percolation problem is
a topic of enormous interest by itself, due to their
applications not only in statistical physics but also in
many other areas such as the study of disordered media,
fluids in porous materials, systems of biological
and ecological interest, etc. \cite{hav1,hav2,stau}.
 , a great progress in the field of the
statistical physics of far-from equilibrium processes
could be achieved by establishing links between
RSA and percolation \cite{Galam,Pekalski,frede}.

Percolation is essentially a geometrical critical
phenomena. The percolation transition is related to the
probability of occurrence of an infinite connectivity between
randomly deposited objects, as a function of the fraction
$p$ of the substrate occupied by the objects \cite{note}.
At the critical point ($p_{c}$), the percolation cluster
in $d-$dimensions,
is a random fractal of dimension $D_{F} = d - \beta/\nu$,
where $\beta$ and $\nu$ are the order parameter and
correlation length critical exponents, respectively.
Close to criticality, the probability $P$ to find a
percolating cluster, on a finite sample of side $L$
can be described by means of an error function \cite{stau}
\begin{equation}
P = \frac{1}{\sqrt{2\pi}\Delta} \int_{-\infty}^{p}
exp\Big{[}-\frac{1}{2}
{\Big{(}\frac{p^{*} - p_{c}}{\Delta}\Big{)}}^2
\Big{]}dp^{*}  ,
\label{error}
\end{equation}
\noindent where $\Delta$ is the width of the transition
region. It is well known that the width vanishes in the
thermodynamic limit according  to \cite{stau}
\begin{equation}
\Delta \propto L^{-\frac{1}{\nu}}.
\label{delta}
\end{equation}
Eq. (\ref{delta}) is very useful because it allows
for the measure of the correlation length exponent $\nu$
that governs the divergence of the correlation length
as $\xi \propto |p - p_{c}|^{-\nu}$ (for random percolation
with $d=2$ one has $\nu = 4/3$).

Very recently it has been suggested that the jamming
probability and the fluctuations of the jamming coverage
may obey relationships similar to Eqs. (\ref{error}) and
(\ref{delta}) \cite{Galam}, respectively. Also, for the
jamming upon RSA of needles in two dimensions the value
$\nu_{J} = 1.0 \pm 0.1$ has been reported \cite{Galam} and this
figure is independent of the aspect ratio of the needles.
Furthermore, early numerical results of Nakamura for the 
RSA of square blocks
are also consistent with $\nu_{J} \simeq 1$ \cite{Naka}, while
Kondrat and Pekalski \cite{Pekalski} have 
reported $\nu_{J} = 1.00 \pm 0.05$ for the RSA of 
segments on the square lattice. Since the obtained values for the
exponent are independent (within error bars) of: i) the
length of the segments (for all a = 1,2,....,45) \cite{Pekalski},
ii) the aspect ratio of the needles \cite{Galam} and iii) 
the size of the square blocks \cite{Naka},
it has been suggested that $\nu_{J}$ is a good candidate for 
an universal quantity of the jamming process \cite{Pekalski}.

The aim of this manuscript is to provide a 
qualitative derivation of Eq. (\ref{delta}) for 
the case of RSA on heterogeneous media. 
It is shown that the exponent
$\nu_{J}$ can be obtained as a function of the 
dimensionality $D$ of the space and the 
fractal dimension $d_{f}$ of the
subset of sites where the RSA process actually takes place.
Our main result
\begin{equation}
\nu_{J}= {\frac{2}{2D-d_{f}}},
\label{nu} 
\end{equation}
\noindent provides a solid ground to previous numerical
data \cite{Galam,Pekalski,Naka}. In fact, these data were obtained 
in $D = 2$ and $d_{f} = 2$, so it follows straightforwardly from
Eq. (\ref{nu}) that $\nu_{J} = 1$ exactly.
Furthermore, in this work, the validity of the proposed relationship 
(\ref{nu}) is verified by means of extensive numerical simulations, 
using both homogeneous substrates as well as different  
random fractals. 

\section{Definitions and Approaches}

When studying RSA on homogeneous samples the definition of the 
jamming coverage and its fluctuations is straightforward, since
one has to deal with a single stochastic process. However,
RSA on nonhomogeneous substrates may
involve the treatment of at least two correlated stochastic processes:
the selection of the particular substrate where the 
deposition is going to take place, and
the RSA process itself. In relation with the first process we will define the 
{\it substrate system} as the set $\{ A_{\lambda}\}$,
composed of $M$ different independent
substrates labeled by the index
$\lambda = 1,...,M$. We consider that $n$ independent RSA processes 
are performed for each element
$A_{\lambda}$ of the substrate system until the jammed state is reached.
We call $ B_{k}^{(\lambda)}$ ($k=1,...,n$) the $n$
different configurations adopted by the entities adsorbed on top of the
substrate. 
The set $\{ B_{k}^{(\lambda)}\}$ taken for the
values of $k$ (for each and all substrate) will be 
referred to as the RSA system.

The jamming coverage $\theta_J$ is a relevant intensive 
quantity that takes the value $\theta_k^{(\lambda)}$ 
when evaluated over the configuration
$B_{k}^{(\lambda)}$. Let us explicitly consider two contributions  to 
$\theta_J$ given by 

\begin{equation}
\theta_J=\theta_{\{sub\}}+\delta\theta_{\{Rsa\}},
\label{suma}
\end{equation}
\noindent It is also assumed that the first term gives 
the most important contribution to $\theta_J$, that can be evaluated according to 

\begin{equation}
\theta^{(\lambda)}= \sum_{k=1}^{n}\frac{\theta_{k}^{(\lambda)}}{n}.
\label{thetalamb}
\end{equation}

\noindent Then, the fluctuations of this term are given by 

\begin{equation}
\sigma_{\{sub\}} =\sqrt{  
\sum_{\lambda=1}^{M} 
\frac{
{({\theta^{(\lambda)}}- \langle \theta_J \rangle)}^2 }
{M-1}}.
\label{sigmasus}
\end{equation}

\noindent where $\langle \theta_J \rangle$ is the average value of 
$\theta_J$, taken over all measurements and configurations

\begin{equation}
\langle \theta_J \rangle= \sum_{\lambda=1}^{M}\frac
{\theta^{(\lambda)}}{M}.
\label{thetamedio}
\end{equation}

The second term in Eq. (\ref{suma})
has a zero average for a fixed substrate.
Some fluctuations in $\theta$ must
also appear from this term and can be characterized by the following average 
\cite{apu1}

\begin{equation}
\sigma^2_{\{ Rsa \}}=
\sum_{\lambda=1}^{M} 
\frac { {\sigma^{(\lambda)}_{\theta}}^2
}
{M} ,
\label{sigmarsa}
\end{equation}

\noindent $\sigma_{\theta}^{(\lambda)}$ being given by

\begin{equation}
\sigma_{\theta}^{(\lambda)}=
\sqrt{ 
\sum_{k=1}^{n}
\frac {{ ({\theta_{k}^{(\lambda)}}-\theta^{(\lambda)}) }^2 }{n-1} } 
\label{sigmarsap}
\end{equation}

It should be stressed that $\sigma_{\{sub\}}$ and
$\sigma_{\{Rsa\}}$ account for the fluctuations on
the substrate and RSA system, respectively.
So, it is expected that these quantities will describe the relevant 
physical behavior related with both sources of randomness.
 
Finally, the total root mean square deviation (RMS) $\sigma_{\theta}$, 
that is expected to describe the fluctuation of the whole system, is given by

\begin{equation}
\sigma_{\theta} = \sqrt {
{\sum_{\lambda=1,k=1}^{M,n}\frac{{
({\theta^{(\lambda)}_k}- \langle \theta \rangle ) }^2 }{Mn-1}}
}
\label{sigmaAB}
\end{equation}

In general, different RMS's are related.
In the simplest case when both contributions in Eq. (\ref{suma}) are
statistically independent, the total fluctuation of $\theta$ is given by

\begin{equation}
\sigma_{\theta} = \sqrt{\sigma_{\{Sub\}}^2+\sigma_{\{Rsa\}}^2}.
\label{sigmaregla}
\end{equation}

As will be shown later, Eq. (\ref{sigmaregla}) is essential for the analysis of
numerical data obtained using finite systems and their corresponding
extrapolation to the thermodynamic limit, because each term may behave
according to different laws. Therefore, simply
measuring ${\sigma_{\theta}}$,
one may obtain inaccurate results, undesired crossover effects and
consequently misleading physical interpretations.

Let us now evaluate the fluctuations associated with the RSA system
using the definition given by Eq. (\ref{sigmarsa}). Let
$n_i = 0, 1$ define the
occupational state of the site $i$.
$N = \sum_{i=1}^{L^D}n_i$ is the number of particles deposited
in a volume $\Omega = L^D$ of a $D$-dimensional 
lattice with periodic boundary conditions. The two-point correlation function
for a homogeneous substrate is given by 

\begin{equation}
G(i,j)=\langle n_in_j \rangle-\langle n_i\rangle \langle n_j \rangle
\label{correfun}
\end{equation}

\noindent where $i,j$ labels sites and $\langle\rangle$ means average
taken over the RSA system in which $N$ varies. 
There is a fundamental relationship between $G$ and the
fluctuation in the number of particles given by

\begin{equation}
\sigma_{N}^2 =  \sum_{i,j=1}^{L^{D}}
G(i,j). 
\label{pura}
\end{equation}

For the case of a unique homogeneous substrate it is well 
known that, when the correlation length
is much shorter than the size of the system, one has

\begin{equation}
\sigma_{N}^2 =  g N_{A}.
\label{pura1}
\end{equation}

\noindent Here $g=\sum_{j=1}^{L^{D}}G(i,j)$ is independent of $N_{A}$,
the number of sites were RSA takes place. The fluctuation of the density
($\theta $) in the thermodynamic limit can be obtained after
dividing both sides of Eq. (\ref{pura1}) by a $L^{2D}$ so that

\begin{equation}
\sigma_{ \{ Rsa \}} \propto  L^{-\frac{D}{2}} ,
\label{homogeneo}
\end{equation}

Let us now generalize this result in order to show so that it still holds for
substrates having sites where deposition is forbidden.  The existence of
these blocked sites on the adsorbing surface breaks the translation
symmetry of the substrate, which was a necessary condition in the above
deduction. However, an equation analogous to Eq. (\ref{homogeneo}) can
still be found if we restrict ourselves to those cases where although each
substrate $A_\lambda$ is nonhomogeneous, the set $\{A_\lambda\}$ is
homogeneous as a whole, i.e., there is no preferred sites on the average.
Stated in a more quantitative way, the following calculations are valid
for systems where the average of $n_i$ over substrates is independent of
$i$.  Then, using a procedure analogous to that employed in the derivation
of Eq. (\ref{sigmarsa}), one can generalize  Eq. (\ref{pura}) by writing

\begin{equation}
\sigma_{N\{Rsa\}}^2 =  \sum_{\lambda=1}^{M}\sum_{i,j=1}^{L^{D}}
\frac{G_{\lambda}(i,j)}{M} = \sum_{i,j=1}^{L^{D}}
(\sum_{\lambda=1}^{M}  \frac{G_{\lambda}(i,j)}{M})
\label{enroque}
\end{equation}

\noindent where the expression within parenthesis could be a measure 
of correlations over the substrate system. However, for a fixed site $i$ 
there may exist substrates in the adsorptive matrix where the deposition 
is forbidden, so that $G_{\lambda}(i,j) \equiv 0$  for all $j$.
Then, since not all the $M$ substrates would be contributing in the sum 
over $\lambda$, simply dividing by $M$ one cannot obtain the real average 
of $G$ over $\lambda$. In order to have a proper average of the RSA correlation 
function, it is necessary to  introduce a factor $X_i$, such that $X_iM$ 
is the actual number of substrates with $G_{\lambda}(i,j) \neq 0$.
It should be noticed that for systems that are homogeneous as a whole 
one has $X_i \equiv X$, so that

\begin{equation}
\sigma_{N\{Rsa\}}^2 = X\sum_{i,j=1}^{L^{D}}
\sum_{\lambda=1}^{M} \frac{G_{\lambda}(i,j)}{MX} =  X\sum_{i,j=1}^{L^{D}}
  G^*(i,j).
\label{uno}
\end{equation}

In order to evaluate $X$, the fraction of substrates 
where a site is not blocked, one has to count the 
number of non-zero terms in Eq. (\ref{enroque}). Adding first over $j$ one 
obtains 

\begin{equation}
g_{\lambda}(i) = \sum_{j=1}^{L^{D}}G_{\lambda}(i,j) .
\label{gchica}
\end{equation}

\noindent Notice that this procedure implies that 
$g_{\lambda}(i) \equiv 0$ for blocked $i$-sites
in a fixed substrate $\lambda$, while $g_{\lambda}(i) \neq 0$ otherwise. 
The remaining double sum 

\begin{equation}
\sum_{\lambda=1,i=1}^{M,L^{D}} g_{\lambda}(i) ,
\label{sumaboba}
\end{equation}

\noindent can be computed fixing $\lambda$ and running $i$ 
over the sites. Then one obtains that the number of contributing terms
is exactly the number of active sites in this substrate. Therefore, the
summation over $\lambda$ has $ML^{d_f}$ terms, 
where $L^{d_{f}}$ is the average number of active sites of the lattice 
where RSA actually takes place. 

When $i$ is fixed and the sum given by Eq. (\ref{sumaboba}) runs over $\lambda$, 
it yields $MX_i$ terms (instead of $M$). Then the summation
$\sum_{i=1}^{L^D} MX_i$ has $M L^D X_i$ terms. Comparing now both results 
and using again the fact that there is no preferred sites in the global system, 
it follows that $X  = \frac{L^{d_{f}}}{L^D}$.  It should be noticed that this correction 
factor is not totally unexpected. In fact, for a substrate system that is homogeneous 
as a whole, one expects that the fraction of substrates with a fixed site $i$ 
active $X_i$, should equal the average proportion of active sites given by $L^{d_f}/L^D$.

Now, replacing $X$ in Eq. (\ref{uno}), it follows 

\begin{equation}
\sigma_{N\{Rsa\}}^2
 = \frac{L^{d_{f}}}{L^D}\sum_{i,j=1}^{L^{D}}  G^* (i,j).
\label{prince}
\end{equation}

\noindent If the substrate system is homogeneous as a whole, then $G^*$ should 
be invariant under translations, because the average of $G$ over $\lambda$ 
should have the same symmetries than those of the substrate system. 
Thus, $g^*(i) = \sum_{j=1}^{L^{D}} G^*(i,j)$ must be site-independent
(i.e., $g^*(i)= g^*_0$), and then Eq. (\ref{prince}) becomes

\begin{equation}
\sigma_{N\{Rsa\}}^2 = L^{d_f} g^*_0. 
\label{g13} 
\end{equation}

\noindent Finally, if the correlation length $\xi_{\{Rsa\}}^*$
associated with $G^*$ for the RSA process 
is short enough ($\xi^*_{\{Rsa\}} << L$), $g_0^*$ is $L$-independent.
Then the fluctuation of the density ($\theta$) can be obtained 
from Eq. (\ref{g13}) dividing by $L^{2D}$, so that 

\begin{equation}
\sigma_{ \{ Rsa \}} \propto  L^{-\frac{1}{\nu}} ,
\label{g141}
\end{equation}

\noindent where

\begin{equation}
\quad \nu= {\frac{2}{2D-d_{f}}}. 
\label{g14}
\end{equation}

It should be stressed that Eqs. (\ref{g141}) and (\ref{g14}) are 
quite general relationships valid for substrate systems that are homogeneous 
as a whole. Furthermore, the condition that the correlation length of the RSA process 
should be smaller than the system size is usually valid for jammed states, 
where the correlation length is very short.

Eqs. (\ref{g141}) and (\ref{g14}) also holds for homogeneous 
substrates, with $L^{d_f}=L^D$ and  $G_{\lambda} \equiv G$, so $\nu= {\frac{2}{D}}$.
These equations are also valid for nonhomogeneous random substrates where the
dimensionality $D$ of the space may be different to the dimensionality 
of the subset of sites where the RSA process actually takes place.
It is also very interesting to notice that, using these relationships it may  
be possible to evaluate $d_f$ performing RSA both in numerical simulations and 
actual experiments. Furthermore, existing numerical simulations
performed in $D = 2$ dimensions with $d_f = 2$ are
in excellent agreement with Eqs. (\ref{g141}) and (\ref{g14})
(notice that for these conditions it follows straightforwardly from
Eq. (\ref{g14}) that $\nu = 1$ exactly): i)
Nakamura \cite{Naka} has reported $\nu_{J} \simeq 1$
(RSA of square blocks), ii) Vandewalle \cite{Galam} 
have reported $\nu_{J} = 1.0 \pm 0.1$ (RSA of needles), and iii)
Kondrat and Pekalski \cite{Pekalski} have 
reported $\nu_{J} = 1.00 \pm 0.05$ (RSA of segments).

In addition to these promising results, we will provide more
astringent tests of the validity of Eq. (\ref{g14}) performing 
numerical simulations of the RSA of dimers on homogeneous substrates 
in $D = 1, 2$ dimensions and as well as using fractal substrates.

\section{Details on the Numerical Simulation of RSA of Dimers}

\subsection{RSA of dimers on Binary alloys}
The first set of simulations are performed for the RSA of dimers 
(the RSA system being the adsorbed atoms) on a binary alloy ({\bf BA}) annealed
at a given temperature (substrate system). 
The {\bf BA} is simulated using the isomorphism with the
Ising model \cite{Binder}, namely spin-up $\equiv A$-species 
and spin-down $\equiv B$-species.
The square lattice of side $L$ in $D = 1,2$ dimensions with
nearest-neighbor (NN) interactions will be considered. 
The {\bf BA} is in contact with
a thermal bath at temperature $T$. The system is assumed to obey
Kawasaki dynamics \cite{Binder}, so that the density of species
$A$ and $B$ is conserved with $\rho_A=\rho_B=1/2$.
The Hamiltonian ($H$) is given by
\begin{equation}
H=E_0-J\sum_{\langle i,j \rangle}s_is_j 
\label{hamilt}
\end{equation}
\noindent where $E_0=N(\epsilon_{AA}+\epsilon_{BB}+2\epsilon_{AB})$, 
$J=-(1/4)(\epsilon_{AA}+\epsilon_{BB}-2\epsilon_{AB})$, 
$\epsilon_{XY}$ is the interaction energy between $X$- and $Y$-species, 
and $s_i=\pm 1$ indicates $A,B$ sites so that $\sum_{i} s_i=0$ \cite{yeoman}.
The Monte Carlo time step (MCS) involves $L^D$ trials, such that each species
of the sample is selected once on average. Stationary
configurations of the {\bf BA} are obtained after disregarding
$10^5$ MCS. It is well known that for $D = 1$ the system is not
critical, exhibiting a (trivially) ordered phase only for $T = 0$.
However, in $D = 2$ dimensions the system undergoes an 
order-disorder transition at $T_{c} = 2.269... $, where $T_{c}$ is 
the Onsager critical temperature (the temperature $T$ is measured in 
units of $J$ setting Boltzmann constant at unity). 
Below this transition temperature, the dynamics of the underlying 
Ising model can be very slow because the system gets trapped on different 
metastable states separated between each other
by high energy barriers. In order to have meaningful 
and systematic results, we have taken the same initial configuration 
for the simulation of each low temperature substrate ($T<T_c$). 
The starting configuration chosen was the one with lowest energy, 
in which $A$- and $B$-species are segregated into two
identical domains, separated by non-defective (straight) domain-walls. 
 
RSA of dimers on the {\bf BA} is assumed to take place on top of NN sites occupied 
by unlike species, i.e., $AB$-sites, adsorption on
$AA$- and $BB$-sites being forbidden. Also multiple occupation
of $AB$-sites is forbidden. These assumptions are based on the
dissociative chemisorption of diatomic molecules ($O_2, H_2,
N_2,$ etc.) on binary catalysts. It is well known that 
in the absence of dimer's diffusion there is a jammed state
\cite{privman}, so that the relevant quantity is the jamming coverage ($\theta_{J}$).
In the case of homogeneous ($D=1$) lattice one 
has the exact result $\theta_J =1-e^{-2}\simeq 0.86466472$ \cite{Evans}.
Also, in the homogeneous ($D=2$) lattice one has $\theta_J \approx 0.906$ \cite{Evans}.
For the present case of a {\bf BA} one expects the coverage to
depend on the temperature at which the substrate has been annealed. Also, it should
satisfy $\theta^{(AB)}_J \leq \theta_J$,
due to the additional constraint
that dimmers can only be adsorbed on specific pairs of sites of
the lattice.

\subsection{RSA of dimers on fractal surfaces}
We have also studied the RSA of
dimers on
stochastic fractals such as the diffusion front \cite{hav1,hav2}
and Ising clusters \cite{Binder}. It is important to 
stress that these chosen fractals satisfy the 
required property of invariance due to their stochastic nature, so that
Eqs. (\ref{g141}) and (\ref{g14}) are expected to hold.

Considering a {\bf BA} at $T=T_C$, the greatest cluster made up taking NN 
sites occupied by the same species, is selected for different lattice sizes. 
This kind of substrate, called spin clusters of the Ising model (SCIM), are fractals 
with a fractal dimension given by
\begin{equation}
d_f= D-\frac{\beta}{\nu} ,
\end{equation}
\noindent where $D$ is the Euclidean dimension, $\beta$ and $\nu$ are the
order-parameter and the correlation-length critical exponents, respectively.
For $D=2$ one has $\beta=\frac{1}{8}$ and  $\nu=1$, so that $d_f=15/8$.

In order to simulate the RSA of dimers on SCIM's,  a site of the cluster 
is selected at random. If that site is empty, a NN site is also selected at
random. The adsorption trial is successful only if that second selected 
site is also empty, otherwise if that site is either occupied 
or lies outside the fractal, the trial is disregarded. Therefore,
double occupation of sites belonging to the fractal is forbidden.

RSA along the perimeter of SCIM's has also been studied. For this
purpose, only adsorption events of dimers taking place on two NN sites, such us 
one of then belongs to the cluster and the remaining one is outside it, 
are considered.

Another type of stochastic fractal used for RSA simulations
is that generated by a diffusion front. In fact, it is well known that 
the properties of the diffusion  front \cite{SRG,SRG1,SRG2,MAR1,MAR2}
are closely related to those of the incipient percolation cluster \cite{hav1,hav2,stau}. 
The front is a (stochastic) self-similar fractal of dimension \cite{ojo1}
$d_{f}^{DF} =  10/7 \approx 1.42857$ \cite{SRG,SRG1,SRG2}. 

In order to obtain diffusion fronts suitable for RSA of dimers,
we have simulated the diffusion of particles at random. Hard-core
interactions, on a $D = 2$ square lattice of size $L \times L$, 
has been considered.
There is a source of particles at the first row of the lattice 
$y = 1, 1 \leq x \leq L$ kept at concentration $p(y) \equiv 1$.
Also, at the last row $y = L, 1 \leq x \leq L$, there is a well
such as $p(L) \equiv 0$. So, there is a concentration gradient
along the source-well direction, while along
the perpendicular {\it x}-direction periodic boundary conditions are
imposed. As the concentration $p(y)$ of particles depends on the position,
decreasing from the source to the well, one actually has a
{\bf gradient percolation} system. The
structure of the diffusion front is identical to the structure of
the hull of the incipient percolation cluster \cite{SRG,PG}. Furthermore,
the concentration of particles at the mean front position $y_f$ is the
same as the percolation threshold $p_c$, so that $p(y_f)=p_c$ \cite{SRG,ZS}.

Considering that particles are connected to their first neighbors while
empty sites are connected to both their first and second neighbors, the
system can conveniently be described by means of the following
geographical analogy: 
The {\bf land} is the set of particles connected
to the source and the {\bf sea} is the set of connected empty sites not
surrounded by land. Also, an {\bf island} is formed by groups of particles
not connected with the land, while {\bf lakes} are formed by connected
empty sites surrounded by land. Finally, the  
{\bf seashore} is the part of the land in contact with the sea that
can be identified with the {\bf diffusion front}, which 
is a {\bf self-similar} fractal.

RSA of dimers on diffusion fronts has been simulated using two rules: 
{\bf Rule I} is that used for the case of RSA on the SCIM perimeter (see above);
while using {\bf Rule II} one only allows the adsorption on sites of the 
diffusion front, disregarding adsorption trials on already occupied
sites of the front and sites outside the fractal.  

\section{Simulation results and discussion}

\subsection{RSA of dimers on Binary Alloys}

\begin{figure}
\centerline{{\epsfysize=8.0cm  \epsffile{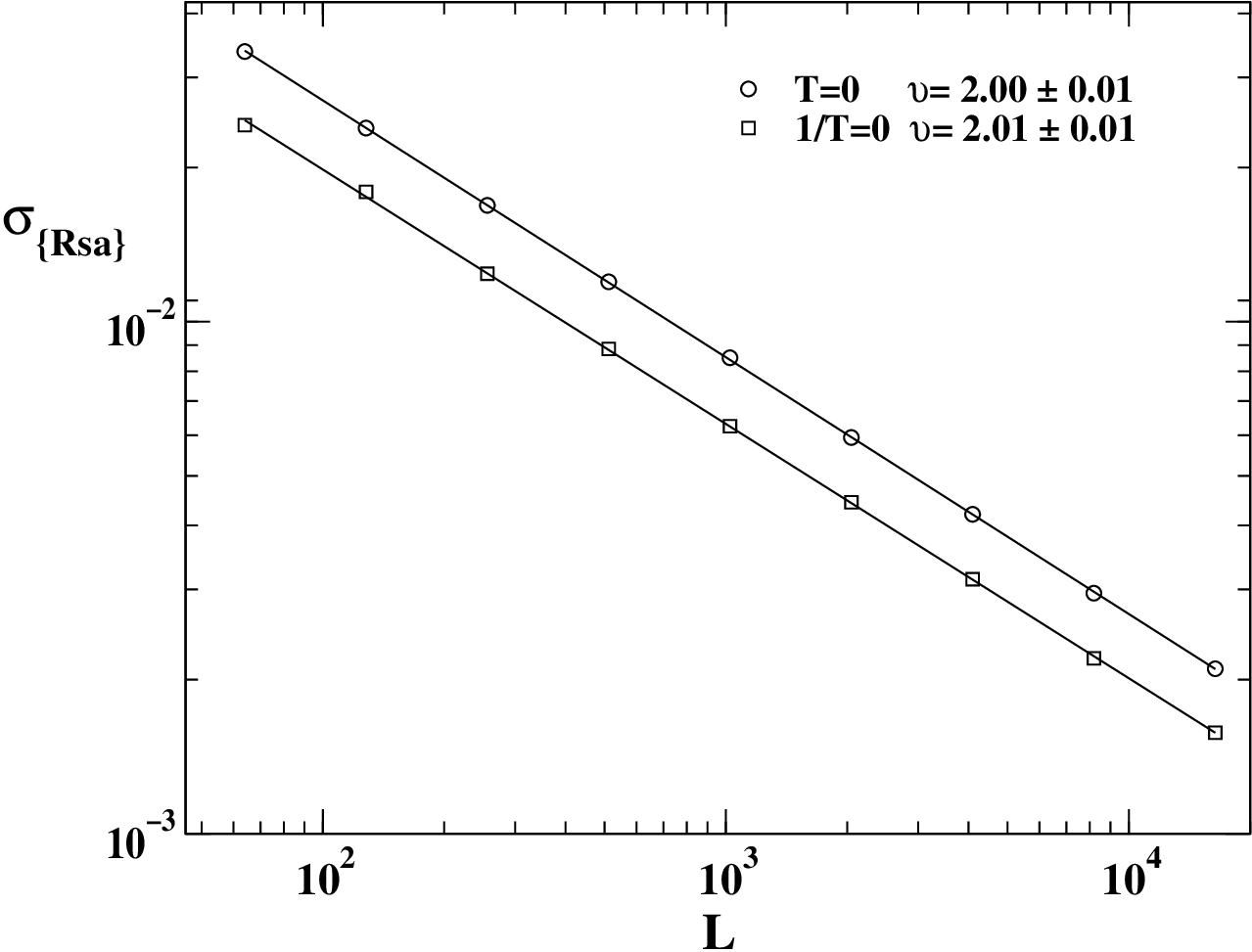}}}
\vspace{0.2cm}
\caption { Log-log plots of $\sigma_{\{Rsa\}}$ versus $L$ 
obtained the case of RSA on {\bf BA}'s in 1 dimension ($D=1$). {\bf BA}'s
at two different temperatures, $T=0$ and $T= \infty$, are considered as 
shown in the figure.} 
\label{BAunaD}
\end{figure}

The simplest test for Eq. (\ref{g14})
corresponds to the case $D = d_f = 1$. Since for {\bf BA} we are only
allowing deposition on unlike sites, this example can be realized 
taking a one dimensional {\bf BA} with nearest-neighbor repulsion
between unlike species (antiferromagnetic Ising system). In fact,
considering the ground state at $T = 0$, corresponding to a fully 
ordered sample, one has that the {\bf BA} is irrelevant and the 
system is equivalent to the standard RSA of dimers in $D= 1$.
On the other hand, for $T > 0$ one has disordered
samples with an homogeneous distribution of blocked sites,
{\it i.e.}, $AA$- and $BB$-sites where dimers are not adsorbed.
Particularly, at $T= \infty$ one has a fully disordered substrate.
The obtained results are shown in Fig. \ref{BAunaD} as log-log plots 
of $\sigma_{\{Rsa\}}$ versus $L$. The best fits of the data give 
$\nu = 2.00 \pm 0.01 $ and $\nu = 2.01 \pm 0.01 $ for the cases 
$T = 0$ and $T = \infty$, respectively. Both figures are in excellent 
agreement with the prediction of Eq. (\ref{g14}) for $D = d_f = 1$, 
namely $\nu = 2$.

\begin{figure}
\centerline{\epsfysize=8.0 cm \epsffile{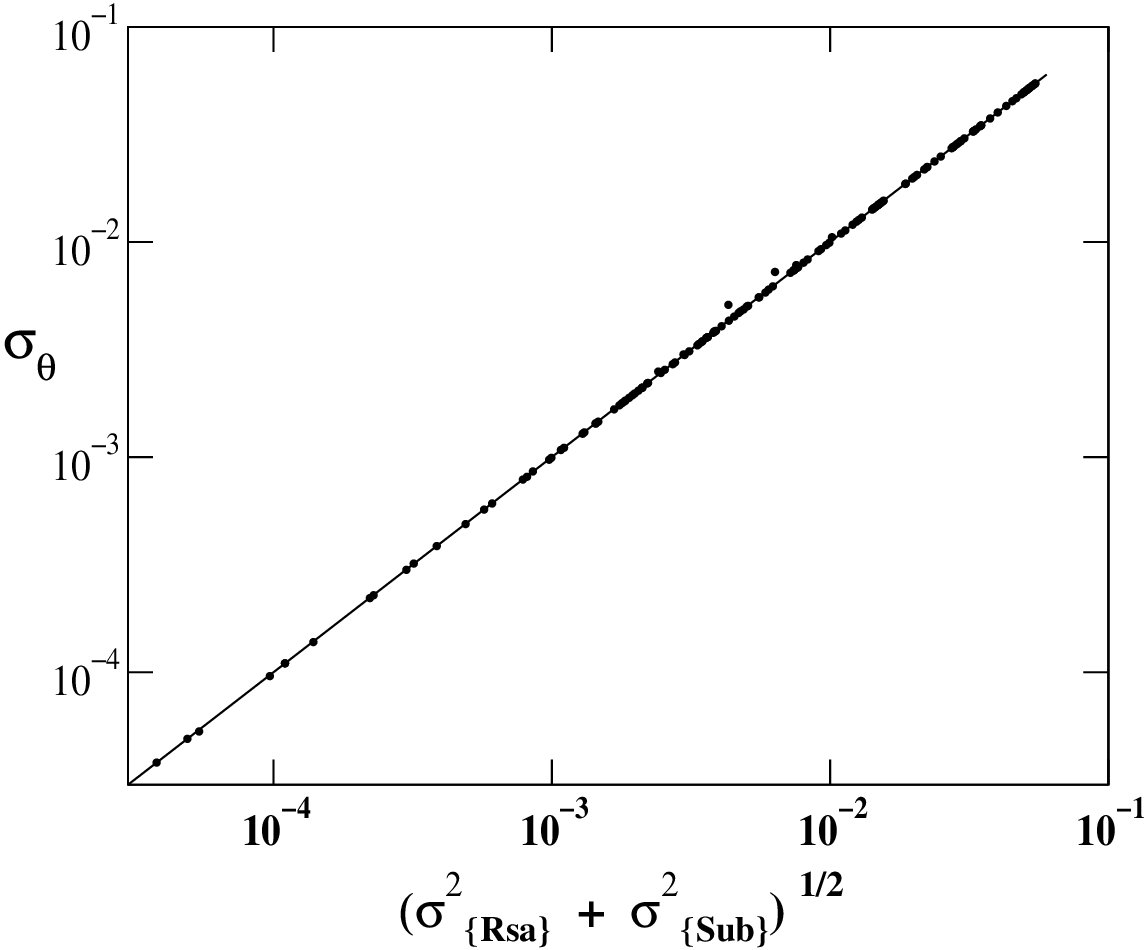}}
\vspace{0.2cm}
\caption { Log-log plots of  $\sigma_{\theta}$ versus  
$\sqrt{\sigma_{\{Sub\}}^2 + \sigma_{\{Rsa\}}^2}$, 
as suggested by Eq. (\ref{sigmaregla}). The different 
terms involved in  Eq. (\ref{sigmaregla}) were obtained according 
to Eqs. (\ref{sigmasus}) and (\ref{sigmarsa}) ({\it horizontal axis})
and (\ref{sigmaAB}) ({\it vertical axis}), respectively . 
The straight line has slope unity.}
\label{BAdosD}
\end{figure}

The second example considered corresponds to RSA on a $D= 2$ {\bf BA} with
NN attractive interactions between alike species,
i.e the ferromagnetic version of the Ising model with conserved
order parameter. In this case one can also tests the validity of 
Eq. (\ref{sigmaregla}) by measuring all the involved terms according to 
Eqs. (\ref{sigmasus}), (\ref{sigmarsa}) and (\ref{sigmaAB}), respectively. 
Fig. \ref{BAdosD} shows log-log plots of 
$\sigma_{\theta}$ versus  $\sqrt{\sigma_{\{Sub\}}^2 + \sigma_{\{Rsa\}}^2}$
obtained  for the case of RSA of dimers on two dimensional {\bf BA}'s.
Data was taken at different temperatures and using lattices 
of different size. The excellent straight line obtained, corresponding to 150 
independent measurements, strongly supports the validity of Eq. (\ref{sigmaregla}).   

\begin{figure}
\centerline{\epsfysize=8.0 cm \epsffile{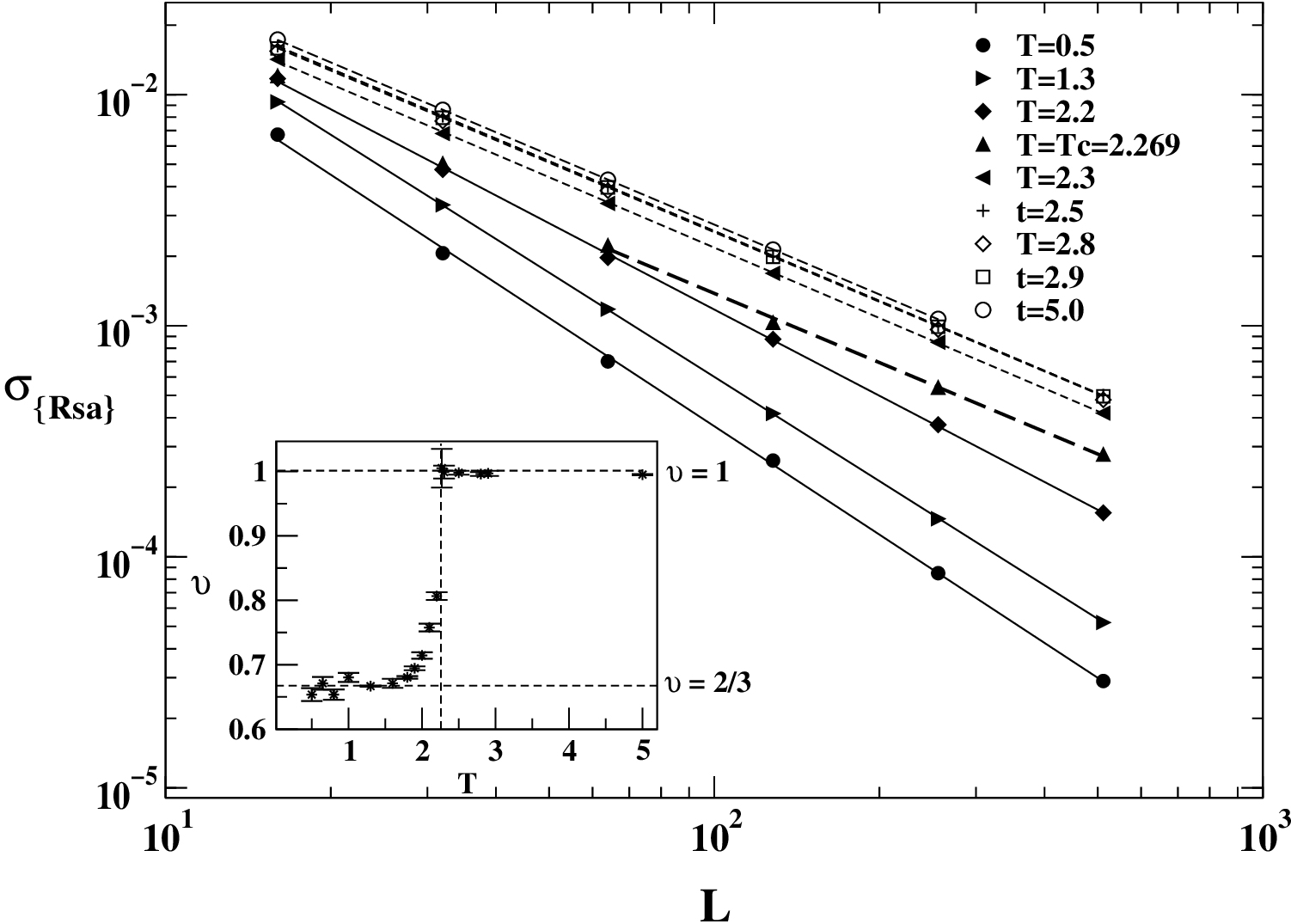}}
\caption { Log-log plots of $\sigma_{\{Rsa\}}$ versus $L$ obtained at different
temperatures performing in each case $n=500$ and $M=500$ measurements.
The inset shows the temperature dependence of the exponent $\nu$
(temperature is measured in units of $J$).}
\label{BAdosDb}
\end{figure}

Also, Fig. \ref{BAdosDb} shows plots of $\sigma_{\theta \{ Rsa \}}$ versus
$L$ for the RSA of dimers on {\bf BA}'s obtained at different temperatures. 
It is found that the power law decay predicted by Eq. (\ref{g141}) always 
holds allowing us to determine the exponents $\nu$.
At low temperatures, below $T_c$,
one has that the {\bf BA} segregates into domains of alike
species containing a certain density of unlike species (impurities)
trapped into the bulk, which increases when the temperature is raised.    
Since dimer adsorption on the bulk of the domains is not possible due to 
the adsorption rule that have been imposed, at very low temperatures the 
RSA process is essentially restricted to the interface between domains. 
In this case one has $d_f = 1$
and Eq. (\ref{g14}) predicts $\nu=2/3$, in excellent agreement
with the results obtained fitting the curves, as shown
in the inset of Fig. \ref{BAdosDb}. Increasing the temperature and 
particularly close to $T_c$, the density of impurities located 
into the bulk of the domains increases. Consequently, an increasing
number of pairs of sites becomes available for the RSA of dimers.   
However, not all these sites contributes to the fluctuations of the
jamming coverage. The simplest example is the case of a single impurity 
surrounded by unlike species where only one  dimer, with four possible
orientations, can be adsorbed. Therefore, fluctuations of the
jamming coverage are only relevant when adsorption takes
place in rather complex arrangements of impurities that are
present in the domains close to $T_c$. For these reasons one
observes a smooth increase of $\nu$ when approaching the 
critical point from below, as shown in the inset of Fig. \ref{BAdosDb}.
Finally, for $T \geq T_C$ one has that adsorption sites, given by
nearest-neighbor pairs of unlike species, are homogeneously distributed 
on the sample with $D=2$ and $d_f=2$. For this case, Eq. (\ref{g14})
predicts $\nu = 1$ in excellent agreement with the numerical results
shown in the inset of Fig. \ref{BAdosDb}. It should be noticed that 
the smooth variation of $\nu$, observed in the inset of Fig. \ref{BAdosDb}
 when approaching $T_c$, may be due to finite-size effects
that hinder the evaluation of the actual exponents. If this is the case,
the exponents $2/3 < \nu_{eff} < 1$, may be considered as effective
size-dependent exponents.

\begin{figure}
\centerline{\epsfysize=7.0 cm \epsffile{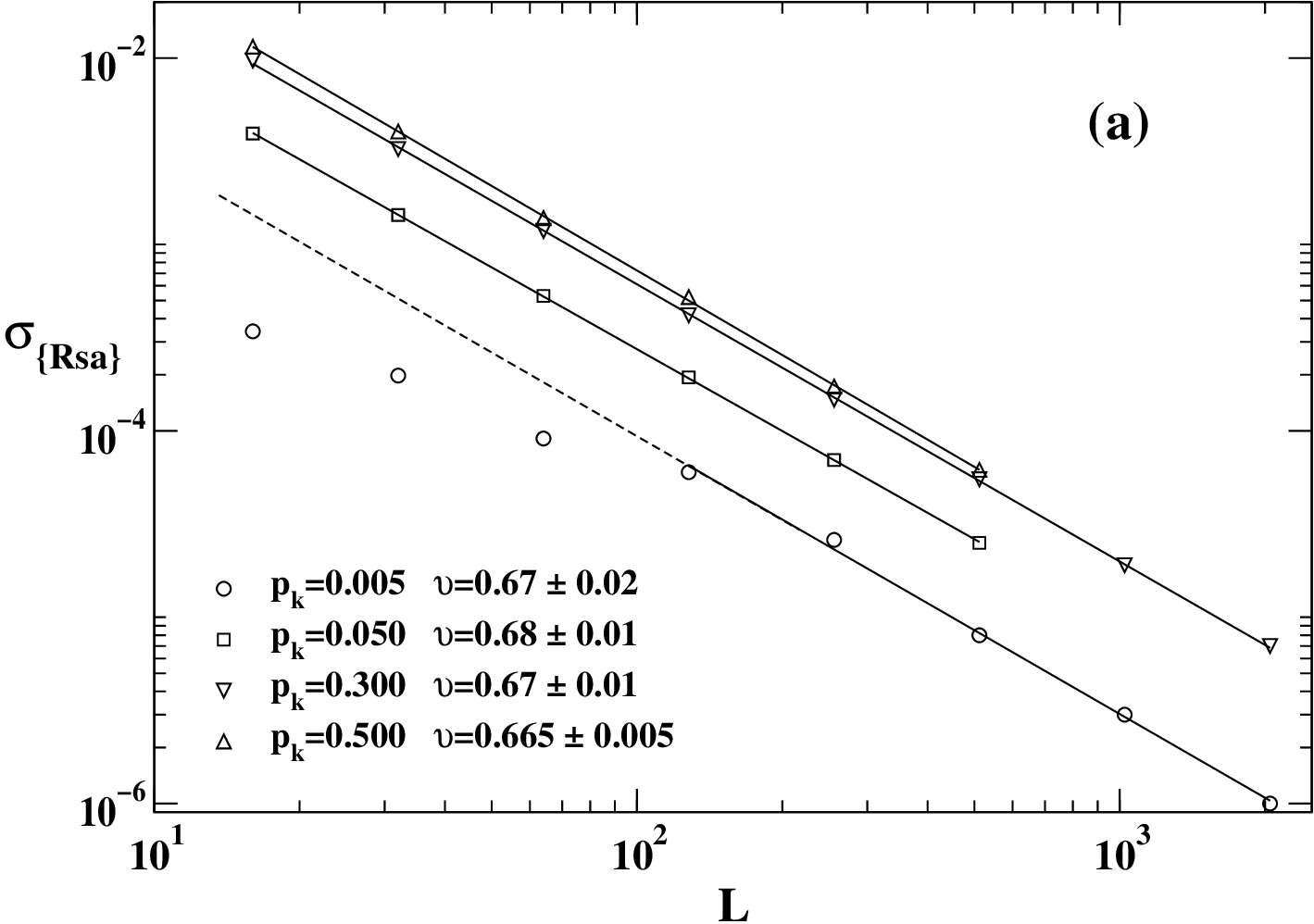}}
\vspace{1.5cm}
\centerline{\epsfysize=8.0 cm \epsffile{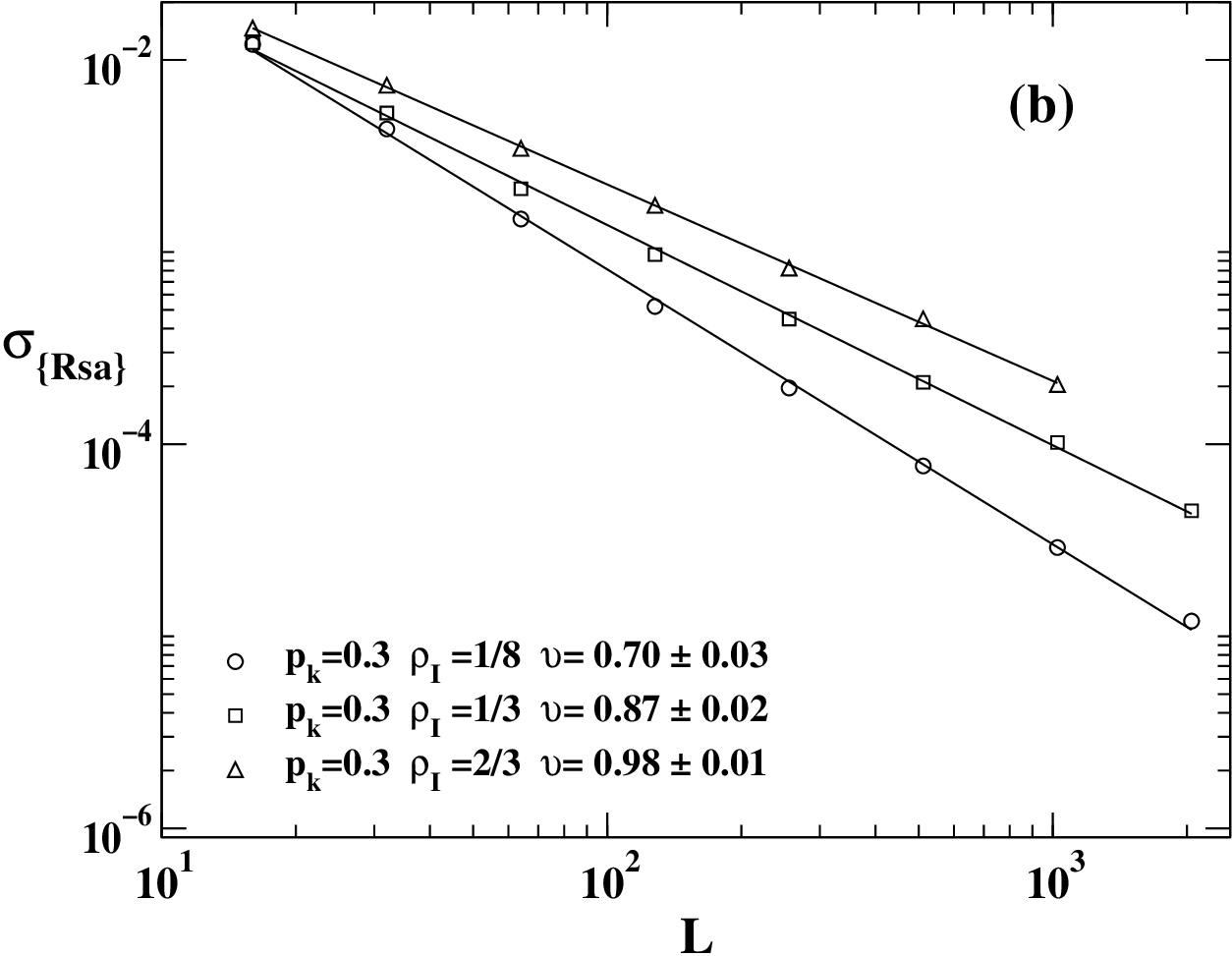}}
\caption {Log-log plots of $\sigma_{\{Rsa\}}$ versus $L$ obtained  
for a simple  one-dimensional interfacial model embedded in a two dimensional
lattice. The parameter $p_k$ allow us to choose different interface roughness,
while $\rho_I$ accounts for the density of  two-dimensional defects in the bulk.
For additional details on the model see the text.
(a) Data corresponding to $\rho_I = 0$ and different values of the roughness,
as listed in the figure. (b) Data obtained keeping the roughness $p_k = 0.3$
fixed and using different values of $\rho_I$, as listed in the figure.}
\label{vivor}
\end{figure}

In order to further support
the above discussed interpretation of the evaluated values of
the exponent $\nu$, numerical simulations using samples having  
controlled interface roughness and concentration of impurities,
have been performed. The starting substrate is a ground state of the
{\bf BA} ($T = 0$) in $D = 2$. The half-left (-right) side of the 
sample is filled with $A$- ($B$-) species, so that a perfectly flat interface
of $AB$-species runs along the middle of the sample. Since RSA on this 
substrate has zero rms, the original interface is modified
introducing kink-like defects with probability $p_{k}$. The resulting 
interface where the RSA process actually occurs is homogeneous
with fractal dimension $d_f = 1$, so that according to Eq. (\ref{g14})
one should expect an exponent $\nu = 2/3$. This prediction is again in 
excellent agreement with the numerical results obtained 
taking $p_{k} = 0.005, 0.05, 0.3$ and $0.5$, as shown in Fig. \ref{vivor}(a).
Of course, for $p_{k} = 0.005$ and small lattices ($L \leq 100$),
finite size effects are observed due to the finite probability
of having a perfectly flat interface without fluctuations.

In addition to the test described previously, samples with a fixed 
interface roughness ($p_{k} = 0.3$) in the example shown 
in Fig. \ref{vivor}(b) were decorated with a controlled density ($\rho_{I}$)
of impurities uniformly distributed in the bulk of the
domains. Fig. \ref{vivor}(b) shows that the presence of impurities not only causes
the fluctuations to increase but also $\nu \rightarrow 1$
for large values of $\rho_{I}$, as expected for $D = 2$ and 
$d_f = 2$ according to Eq. (\ref{g14}). The value
$\nu = 0.70 \pm 0.03$ obtained for $\rho_{I} = 1/8$ is 
very close to the figure $\nu = 2/3$ corresponding to 
$\rho_{I} = 0$ since a majority of isolated and small clusters of 
impurities can not significantly influence neither the 
fluctuations nor the exponent. In fact,
only more complex clusters of impurities, as those formed for 
$\rho_{I} = 1/3$ and $\rho_{I} = 2/3$ in the example shown in
Fig. \ref{vivor}(b), allow for a large variety of adsorption configurations
with an appreciable enhancement of the fluctuations of the 
RSA process that causes a noticeable effect on the exponent.

\subsection{RSA of dimers on fractal surfaces}

\begin{figure}
\centerline{\epsfysize=8.0 cm \epsffile{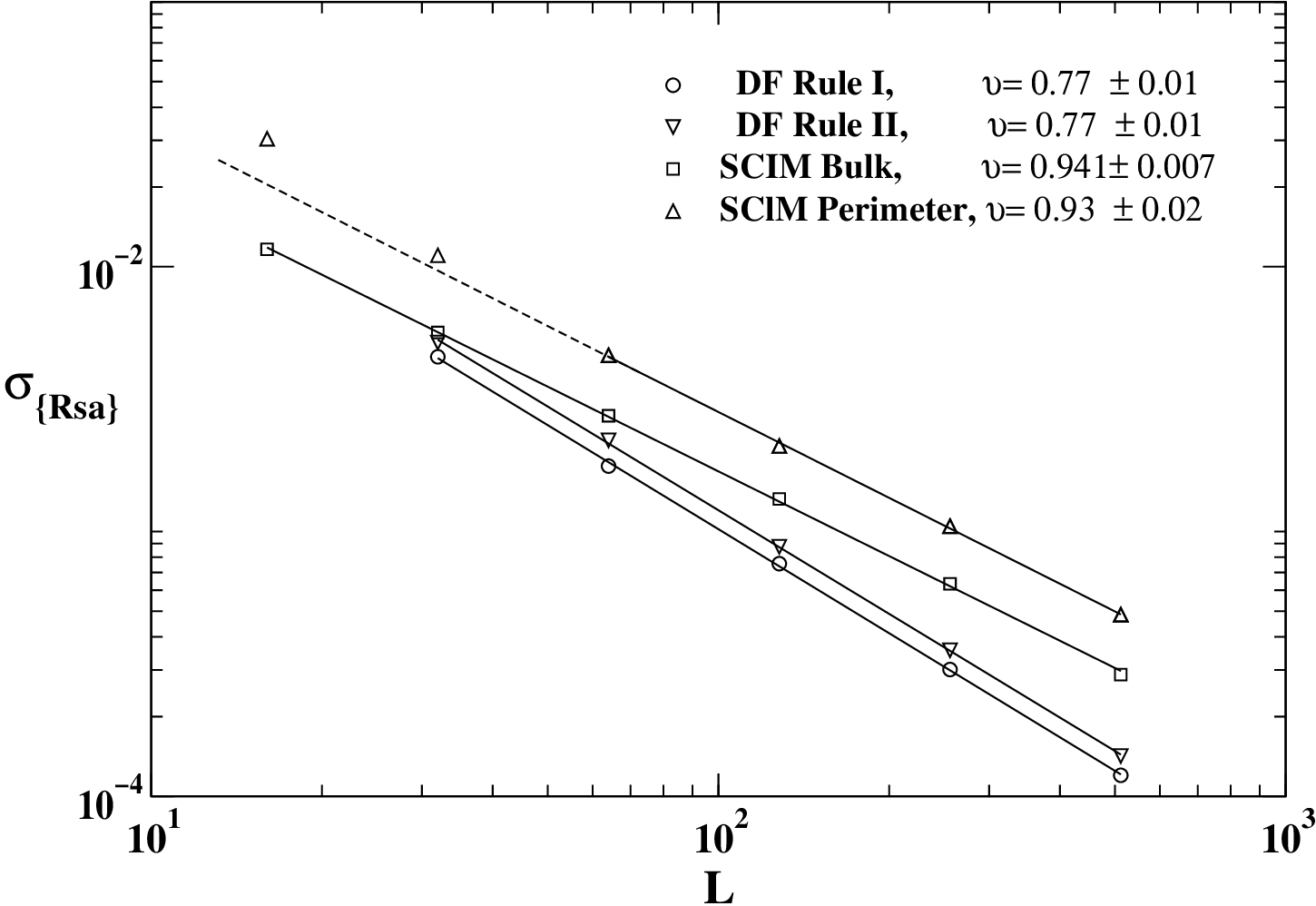}}
\caption {Log-log plots of $\sigma_{\{Rsa\}}$ versus $L$ for the case of RSA on
random fractals. Considering Spin Clusters of the Ising model (SCIM) at $T_c$, 
it is found that Eq. (\ref{g14}) is valid when the RSA process is done on both, 
the bulk and the perimeter cluster. Also, results obtained upon RSA on the 
fractal generated by diffusion fronts (DF) with two different adsorption 
rules, are shown. For details on the SCIM and the adsorption rules see the text.}
\label{randomfractals}
\end{figure}

Fig. \ref{randomfractals} shows log-log plots of $\sigma_{\{Rsa\}}$ versus $L$
obtained upon RSA of dimers on SCIM's. In this example 
one has $D=2$ and $d_f = 15/8$, so that the value
$\nu=16/17 \simeq 0.941176$ is expected according to Eq. (\ref{g14}).
The results obtained fitting the data corresponding to RSA on the bulk of the 
cluster and on its perimeter are $\nu = 0.941 \pm 0.007$ and 
$\nu = 0.93 \pm 0.02$, respectively. In the latter case 
for small lattices ($L \leq 60$), finite size effects are observed 
as in the analogous percolation problem \cite{stau}. Also,
Fig. \ref{randomfractals} shows the results for the RSA of dimers on a
diffusion front in $D = 2$. 
This is an interesting example to test the validity of Eq. 
(\ref{g14}) using an stochastic fractal with a well known fractal 
dimension $d_{f} = 10/7$. For this example one expects 
$\nu =  0.777...$ while the best fit of the numerical data
shown in Fig. \ref{randomfractals} 
gives $\nu = 0.77 \pm 0.01$, using two different adsorption rules as 
discussed above. These results are also in excellent agreement with 
Eq. (\ref{g14}) giving further numerical support to 
the analytical relationship derived in Section {\bf{2}}.

\subsection{Interplay between fluctuations of correlated stochastic processes}
It should be stressed that in order to capture the physics of the 
RSA process, all the fluctuations measured in the previous sections 
were evaluated with the aid of Eqs. (\ref{sigmarsa}) and
(\ref{sigmarsap}).
The aim of this subsection is to show that a quite different physical 
picture can be obtained measuring the fluctuations linked with the 
substrate system. Let us recall that for this purpose one also has to
perform  RSA measurements, however fluctuations have to be evaluated  
using Eqs. (\ref{sigmasus}), which within the context of the present work, 
captures the physics of the underlying substrate where the RSA process takes place. 

As an example, the discussion will be restricted to the case of
RSA of dimers on a {\bf BA} in $D = 2$, in order to perform comparisons with
the data obtained in the simulations shown in Figs. \ref{BAdosD} and 
\ref{BAdosDb}. Considering a single configuration of the alloy and according to
the adsorption rules, one has that $\theta_J$ is essentially a 
measure of the density of $AB$-sites present in the configuration. 
Furthermore, such pairs (which are known as the `broken bonds' in Ising 
spin language), contribute to the internal energy per site of the {\bf BA} ($u$). 
Applying the fluctuation-dissipation theorem \cite{Stanley} it is possible to
show that the specific heat of the {\bf BA} 
($C_V={(\frac{\partial u}{\partial T})}_V$) is given by 
the fluctuations of $u$, namely

\begin{equation}
C_V =  \frac{L^2}{T^2} \sum_{\lambda=1}^{M}\frac{(u_{\lambda}-\langle u \rangle)^2}{M-1}= \frac{L^2}{T^2} \sigma_u^2
\end{equation}

\noindent where averages are taken over the set $\{A_{\lambda}\}$ of different
configurations of the {\bf BA}. On the other hand, for the $D=2 $
Ising model one has that 

\begin{equation}
C_{V}(T) \propto |T-T_C|^{\alpha}  ,
\label{lacalor}
\end{equation}

\noindent  with $\alpha \equiv 0$, i.e. a logarithmic divergence at criticality. 
Analogously, one can define an RSA 'susceptibility' $\chi$ for 
the jammed state as 

\begin{equation}
\chi = L^{2}  \sum_{\lambda=1}^{M}\frac{(\theta^{\lambda}_J-\langle \theta_J \rangle)^2}{M-1}=L^2\sigma_{\{Sub\}}^2  , 
\label{chicho}
\end{equation}

\noindent where it is expected that, if the system conformed by the 
deposited particles would follow (or `copy'), in some way, the structure of 
the underlying substrate,  $\chi$ would follow the same behavior that $C_{V}$. 
\begin{figure}
\centerline{{\epsfysize=8.0 cm \epsffile{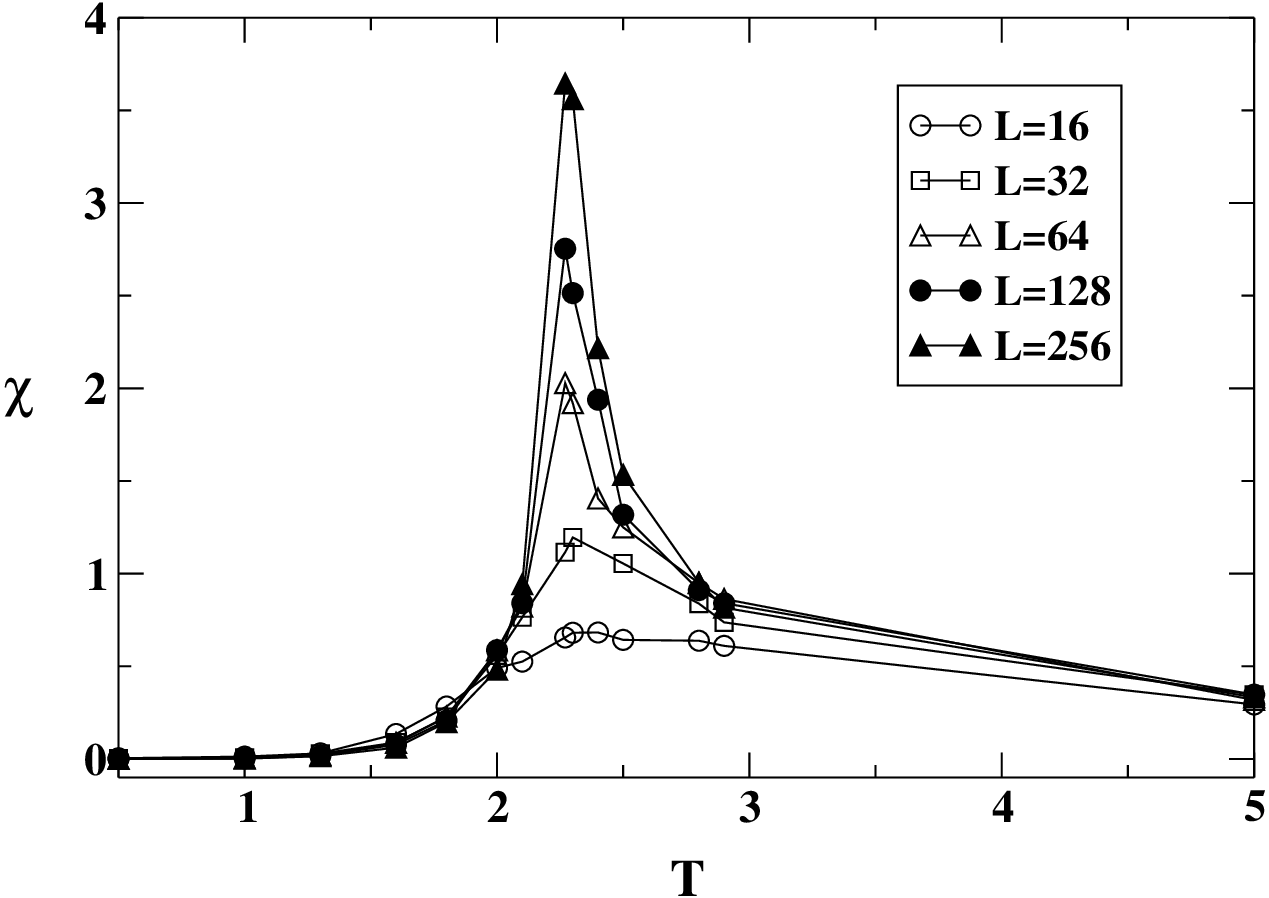}}}
\caption { Plots of $\chi$ versus $T$ obtained for the RSA of dimers on {\bf BA} 
using  lattices of different size $L$ ($T$ in units of $J$). The data exhibits the typical behavior 
characteristic  of a second-order phase transition.}
\label{chiT}
\end{figure}

\begin{figure}
\centerline{\epsfysize=15.0 cm \epsffile{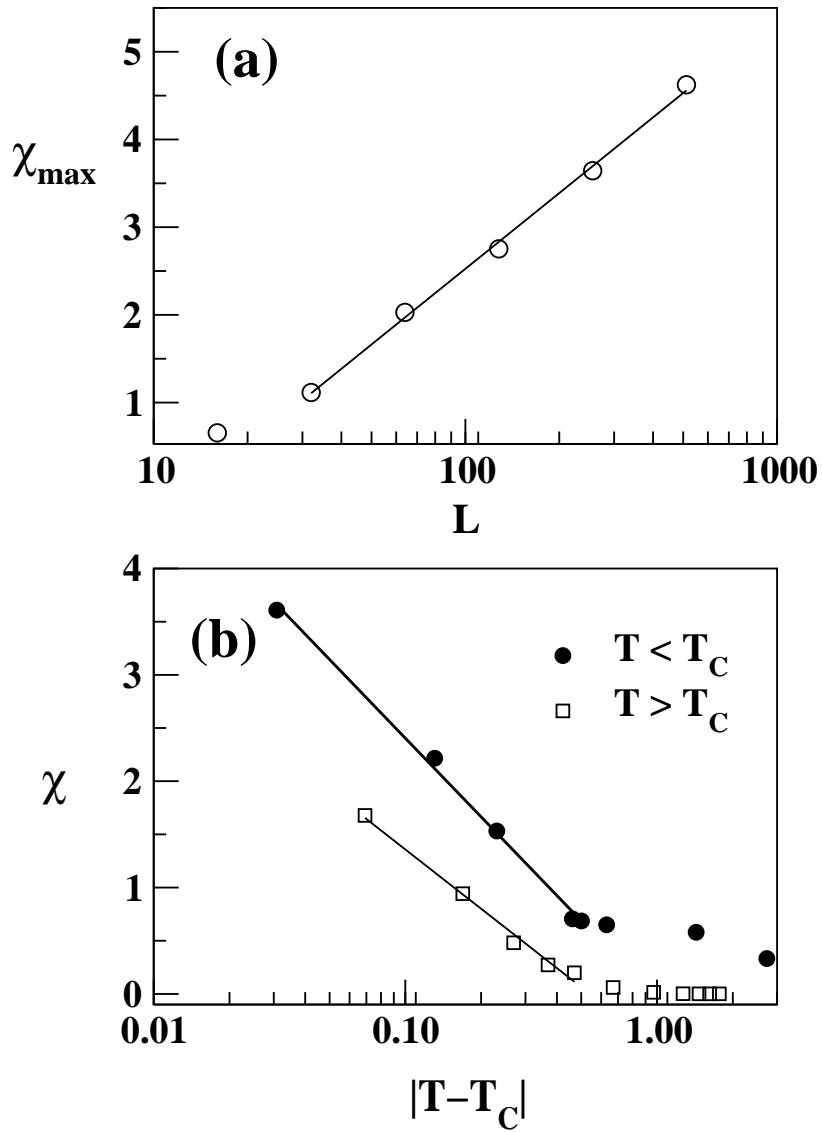}}
\caption {Data corresponding to the RSA of dimers on {\bf BA}'s. 
(a) Linear-log plot of $\chi_{max}$ versus $L$.
(b) Linear-log plot of $\chi$ versus $|T-T_C|$ (in units of $J$) obtained using 
lattices of size $L=256$. Both plots show that $\chi$ reflects
the divergences of the specified heat of the underlying {\bf BA}.}
\label{chidiver}
\end{figure}

Fig. (\ref{chiT}) shows that plots 
of $\chi$ versus $T$, obtained using lattices of different size, 
exhibit clear peaks close to the critical temperature of the underlying {\bf BA},
resembling the behavior of the specific heat in finite samples. In fact,
the peaks are shifted and rounded due to operation of finite-size
effects. This behavior is the typical one for a second-order phase transition 
that implies the existence of a diverging correlation length $\xi$
when approaching criticality according to \cite{Binder}

\begin{equation}
\xi(T) \propto |T-T_C|^{-\nu^{*}}  ,
\label{lacorre}
\end{equation}

\noindent where $\nu^{*} = 1 $ is the correlation length exponent
of the Ising model. Using finite-size scaling arguments
one can set $L^{1/\nu^{*}}|T-T_c| \approx 1$ \cite{Binder}, then  
replacing into Eq. (\ref{lacalor}) with $\alpha = 0$, using
Eq. (\ref{chicho}) and assuming that close to criticality one has 
$\chi \sim  C_{v}$, it follows  

\begin{equation}
\chi_{max}(L) \propto ln(L)  ,
\label{chiL}
\end{equation}

\noindent where $\chi_{max}$ is the maximum value of $\chi$
that can be obtained from the peaks shown in Fig. \ref{chiT}.
The results shown in Figs. \ref{chidiver}(a) and \ref{chidiver}(b)
confirm the divergences of $\chi$  expected according to
Eqs. (\ref{lacalor}) and (\ref{chiL}), respectively.  So, this finding 
shows that there is an additional (divergent) correlation length of the RSA process 
that is associated to the substrate, and it can be captured by measuring $\chi$.  

\begin{figure}
\centerline{\epsfysize=8.0 cm \epsffile{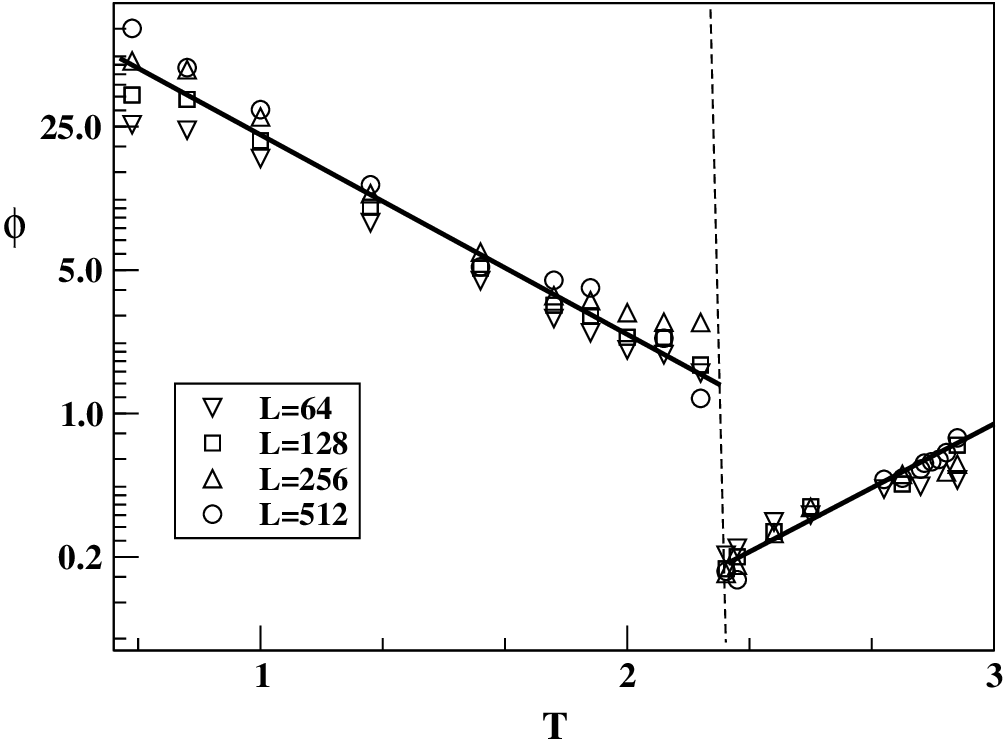}}
\caption { Plot of $\phi$ versus $T$ ($T$ in units of $J$),
 obtained using lattices 
of different size $L$. The jump observed in $\phi$ occurs
at $T_C$ (dashed line). The solid lines have been drawn to guide the eyes.}     
\label{multiescal}
\end{figure}

From Figs. \ref{BAdosDb}, \ref{chiT}  and \ref{chidiver},
as well as according to the above discussion,
it follows that the RSA of dimers on {\bf BA}'s provides an
interesting example of the interplay between two correlated
processes whose respective fluctuations obey
different functions. It is worth mentioning that, in spite of exhibiting quite 
different behavior, both functions capture the criticality of 
the substrate. In fact, the fluctuations of the jammed state, 
as measured according to Eqs. (\ref{sigmasus}) and ({\ref{chicho}), 
reflect the critical behavior of the {\bf BA} through the relationship 
between $\chi$ and $C_v$ while, on the other hand,
using Eqs. (\ref{sigmarsa}) and (\ref{sigmarsap}) the criticality of the {\bf BA} becomes
evident through the jump of the exponent $\nu$ observed close to $T_c$
(see Fig. \ref{BAdosDb}). It is also interesting to remark that the 
relationship that relates the different fluctuations of both processes,
given by Eq. (\ref{sigmaregla}), does not show such a critical behavior at all, 
as shown in Fig. \ref{BAdosD}.
 
In order to further analyze these results, let us define 
$\phi^* = \sigma_{\{Rsa\}}/\sigma_{\{Sub\}}$, 
such as $\phi^*$ gives a measure of the relative intensity of the RSA fluctuations 
as compared to those of the substrate. Then, 
using Eqs. (\ref{chicho}), (\ref{g141}) and (\ref{chiL}), it follows that

\begin{equation}
\phi^{*} = \phi(T) f(L) = \phi(T) \frac {L^{-1/\nu}}{L^{-1}\sqrt{ln(L)}} ,
\label{fi}
\end{equation}

\noindent where the $L$-dependence behavior of $\phi^{*}$ appears explicitly 
through the function $f(L)$, while $\phi(T)$ accounts for the temperature dependence. 
Fig. \ref{multiescal} shows that log-linear plots of $\phi(T) = \phi^{*}/f(L)$ versus
$T$ exhibit an acceptable collapsing. At both sides of $T_C$ the function 
$\phi(T)$ approximately follows a logarithmic behavior, resembling the temperature 
dependence of $\chi$ near $T_c$. Moreover a clear jump appears at $T_{c}$, 
due to the change in the fractal dimension of the adsorbing set of sites of 
the substrate ($d_f=1$ for $T < T_{c}$ and $d_f=2$ for $T \geq T_{c}$)
observed in the finite samples used in the simulations. 
This result implies that the temperature dependence of the structure of the substrate
prevails over that of the RSA process, and that the fluctuations
due to substrate also prevails over those due to the RSA process 
(notice that $f(L) \rightarrow 0$ when $L \rightarrow \infty$). 


\section{Conclusions}

In this paper the behavior of the fluctuations of the jamming coverage
upon RSA process on homogeneous and nonhomogeneous substrates has been studied.
Pointing the attention to the RSA process and applying the definition given
by Eq. (\ref{sigmarsa}) in order to measure the fluctuations, we have shown  
both analytically and by means of numerical simulations,
that the jamming fluctuations behaves as ${\sigma_{\theta_{J}}\propto
L^{- 1 / \nu_{J}} }$ , where $\nu_{J}$ is given by 
$\nu_{J}= {\frac{2}{2D-d_{f}}}$, $D$ and $d_f$ being the dimension of the 
lattice and that of the active sites where adsorption actually takes place, 
respectively. These results are suitable to describe systems characterized by
a short-range correlation length of the RSA process, that  may take place on both,
homogeneous and nonhomogeneous substrate. From our derivation of Eq. (\ref{g14}) 
it follows that $\nu_{J} = 1$ for $D = 2$ and $d_{f} = 2$, in accordance
with results published previously by various authors 
\cite{Galam,Pekalski,Naka}.

It should also be noticed that Eq. (\ref{g14}) allow us to measure the 
fractal dimension of the different adsorption sets where deposition 
takes place according to the specified adsorption rules. Summing up, 
our results not only point out that a careful treatment of the
fluctuations of correlated processes is necessary in order to capture the 
desired physical behavior, but also provide a tool 
for the evaluation of fractal dimensions using RSA experiments.

{\bf Acknowledgments}: This work was supported by CONICET, 
UNLP and ANPCyT (Argentina).


\begin{thebibliography}{99}
\bibitem{6} A. Patrykiejew, S. Sokotowski and K. Binder, 
Surf. Sci. Rep. \textbf{37}, 207 (2000).

\bibitem{7} T. L. Einstein, \emph{Chemistry and Physics of Solid Surfaces}, 
Vol. 4, Ed. R. Vanselow and R. Hove, Springer-Verlag, Berlin (1982).

\bibitem{8} W. H. Weinberg, Ann. Rev. Phys. Chem. \textbf{34}, 217 (1983).

\bibitem{noev} R. S. Nord  and J. W. Evans,
J. Chem. Phys. {\bf 82}, 2795 (1985).

\bibitem{teno} J. t. Terrell and R. S. Nord;
Phys. Rev. A., {\bf 46}, 5260 (1992).

\bibitem{Evans} J. W. Evans, Rev. Mod. Phys., {\bf 65}, 1281 (1993).

\bibitem{9} J. J. Ramsden, J. Stat. Phys. \textbf{73}, 853 (1993).

\bibitem{Galam} N. Vandewalle, S. Galam and M. Kramer, Eur. Phys. J. B, 
{\bf 14}, 407 (2000).

\bibitem{Pekalski} G. Kondrat and A. Pekalski, Phys. Rev. E, 
{\bf 63}, 051108 (2001)

\bibitem{frede} F. Rampf and E. V. Albano, Phys. Rev. E.,{\bf 66}, 061106 (2002).

\bibitem{hav1} {\it Fractals and Disordered Systems}, 
Eds. A. Bunde and S. Havlin. Springer-Verlag.  Heildelberg, (1991).

\bibitem{hav2} {\it Fractals in Science}, 
Eds. A. Bunde and S. Havlin. Springer-Verlag.  Heildelberg, (1994).

\bibitem{stau} D. Stauffer and A. Aharony, {\it Introduction
to the Percolation Theory}, Taylor and Francis, London, (1992) 2nd edition.

\bibitem{note} Notice that the occupation probability
$p$, in the percolation framework, is equivalent
to the coverage $\theta$ used within the
terminology of RSA.

\bibitem{Naka} M. Nakamura, J. Phys. A, {\bf 19}, 2345 (1986).

\bibitem{apu1} Instead of using Eq. (\ref{sigmarsa}) one can also evaluate
the fluctuations as
$\langle \sigma_{\{ Rsa\}} \rangle =\sum_{\lambda=1}^{M} 
\frac {\sigma^{(\lambda)}_{\theta}}{M} $. However, the 
relative difference between both,   
$\sigma_{\{ Rsa \}}$ and $\langle \sigma_{\{ Rsa\}} \rangle$, is negligible.
In fact, evaluating 
$\frac{\sigma_{\{ Rsa \}}} {\langle \sigma_{\{ Rsa\}}\rangle} 
\approx (1+ \frac{1}{2} \epsilon^2)$, one has that   
$\epsilon$ is a second-order correction such as 
$\epsilon =\frac{\sigma_{x}}{x}$ with $x=\langle \sigma_{\{ Rsa\}} \rangle$.
We have checked that in our measurements one allways has  $\epsilon < 0.1$. 

\bibitem{Binder}  {\it The Monte Carlo Method in Condensed Matter Physics},  
Ed. K. Binder. Springer-Verlag, Berlin, (1992).

\bibitem{yeoman} J. M. Yeomans, 
{\it Statistical Mechanics of phase transitions.}  
Clarendon Press, Oxford, (1992).

\bibitem{privman} {\it Nonequilibrium Statistical Mechanics in One
Dimension} edited by V. Privman, Cambridge University Press
Cambridge, 1997 and references therein

\bibitem{ojo1} It can be demonstrated that 
$N_f \propto L^{\frac{2d_o-1}{d_o}}$, where 
$N_f$ is the number of sites belonging to the diffusion front, 
$d_o=\frac{7}{4}$ is its fractal dimension of the inner perimeter of the front,
$L$ is the lateral extension of the front and $L^{-1}$ is the density gradient
along the sample.   

\bibitem{SRG} B. Sapoval, M. Rosso and J. F. Gouyet. 
J. Physique.  Lett. {\bf 46}, L49 (1985).

\bibitem{SRG1} M. Rosso, J. F. Gouyet and B. Sapoval. 
Phys. Rev. B. {\bf 32}, 6053 (1985).

\bibitem{SRG2} M. Rosso, J. F. Gouyet and B. Sapoval. 
Phys. Rev. Lett. {\bf 57}, 3195 (1986).

\bibitem{MAR1} A. Memsouk, Y. Boughaleb, R. Nassif and H. Ennamiri.
Eur. Phys. J. B. {\bf 17}, 137 (2000).

\bibitem{MAR2} A. Hader, A. Memsouk and Y. Boughaleb.
Eur. Phys. J. B. {\bf 17}, 137 (2000).


\bibitem{PG} P. Grasberger. J. Phys. A. (Math. and Gen.). 
{\bf 19}, 2675 (1986).

\bibitem{ZS} R. M. Ziff and B. Sapoval. 
J. Phys. A. (Math. and Gen.). {\bf 19}, L1169 (1986).


\bibitem{Stanley} H. E. Stanley, {\it Introduction to phase transitions 
and critical phenomena.} Oxford University Press, New York, (1971).

\end{thebibliography}
\end{document}